\documentclass[sigconf]{acmart}
\usepackage{tabularx}

\renewcommand\footnotetextcopyrightpermission[1]{} % removes footnote with conference information in first column

\setcopyright{none}

\acmConference[40th Brazilian Symposium on Software Engineering]{40th Brazilian Symposium on Software Engineering}{2026}{São Paulo, SP, Brazil}

\AtBeginDocument{
    
}

\begin{document}

%% allowing the author to define a "short title" to be used on the page 
%% headers.
\title{GADR: Gathering Architecture Decision Records from Meeting Transcriptions}

\author{Lucas Daniel Costa da Silva}
\affiliation{
  \institution{Universidade Federal de Pernambuco (UFPE)}
  \city{Recife}
  \country{Brasil}
}
\email{ldcs@cin.ufpe.br}

\author{Kiev Gama}
\affiliation{
  \institution{Universidade Federal de Pernambuco (UFPE)}
  \city{Recife}
  \country{Brasil}
}
\email{kiev@cin.ufpe.br}

%% By default, the full list of authors will be used on the page
%% headers. This list is often too long and will overlap
%% other information printed in the page headers. 
%% This command allows the author to define a more concise list
\renewcommand{\shortauthors}{Silva and Gama}
\renewcommand{\shorttitle}{GADR: Gathering ADRs from Meeting Transcriptions}

%% The abstract is a short summary of the work the paper presents.
\begin{abstract}
%In Software Engineering, Architectural Knowledge Management (AKM) plays a critical role in modern development practices. Architecture Decision Records (ADRs) serve as a primary instrument for documenting the context, rationale, and alternatives considered during software design decisions. 
Existing LLM-based approaches to Architecture Decision Record (ADR) generation share a critical and largely unexamined assumption: that input is already reasonably structured. In practice, architectural decisions emerge from informal, noisy meetings where choices are implicit, fragmented, and entangled with off-topic dialogue, precisely the conditions under which single-pass prompting degrades. This paper presents GADR, a multi-agent, self-correcting workflow that extracts architectural decisions from raw meeting transcriptions and generates Nygard-formatted ADR drafts. A feasibility study comprising five real project meeting transcripts, expert review by four senior architects, and evaluation by fifteen students provides initial evidence that the agentic workflow captures most expert-identified decisions and produces drafts participants found clear and useful, outperforming zero-shot and few-shot baselines in stability and structural adherence. The study also addresses the underexplored trade-off of RAG-based enrichment improving ADR depth while simultaneously risking transcript-unfaithful content, raising open questions about traceability in automated architectural documentation that we believe is worth the community's attention.
%However, ADR adoption continues to be limited in practice, as the effort required for manual creation disrupts the natural development workflow. Although Large Language Models (LLMs) have demonstrated potential in automating documentation generation, single-pass approaches (such as zero-shot and few-shot prompting) yield unreliable results when processing unstructured, noisy input data, such as meeting transcriptions. To address this limitation, this research proposes an agent-based approach for the automatic generation of ADRs from technical meeting transcriptions. The implemented architecture incorporates a iterative refinement loop designed to mitigate hallucinations and verbosity, while concurrently persisting the generated ADRs in accordance with the most widely adopted community standard (Nygard). An empirical evaluation was conducted in an academic setting, in which students enrolled in the Software Architecture course within a Computer Science program will provide transcriptions of their meetings and complete a survey assessing the generated ADRs in terms of precision and clarity. The evaluation criteria focus on the accuracy of the context and decisions extracted from the meeting transcriptions, ensuring the faithful capture of key architectural decisions discussed during the meetings, thereby enabling the sustainable adoption of ADRs in practice.
\end{abstract}

%% the presented work. Separate the keywords with commas.

%% This command processes the author, affiliation, and title
%% "frenchspacing" avoids an additional space after a period at the end of a sentence.
\frenchspacing
\maketitle

\section{Introduction}
Software Architecture is increasingly defined not just by structural elements but by the decisions that shape them~\cite{Bass2003,Jansen2005}. Architecture Decision Records (ADRs) were proposed as a lightweight mechanism to document those decisions~\cite{Nygard2011,Nogueira2026,keeling2017design}, that emerge as a means to enhance Architectural Knowledge Management (AKM) within projects, ensuring that decisions made by various stakeholders over time can be understood within their historical context. %Yet their adoption in practice remains low, primarily because the effort of capturing architectural knowledge disrupts the development workflow~\cite{Buchgeher2023}.

%A software architecture comprises software elements (i.e., modules), the externally visible properties of those elements, and the relationships among them \cite{Bass2003}. This implies that architecture serves as an abstraction focused on element interfaces, thereby concealing internal implementation details. Design decisions have become an explicit component of the architectural definition, as Jansen and Bosch defined Software Architecture as a composition of a set of Architectural Design Decisions \cite{Jansen2005}. This approach specifically aims to mitigate the vaporization of architectural knowledge.

%In turn, Architecture Decision Records (ADRs) were proposed by Nygard in 2011 as a mechanism to document these architectural decisions \cite{Nygard2011}, characterized by a lightweight approach\cite{Nogueira2026}. An ADR can be authored in any text editor, though Markdown is the preferred format for standardizing Architecture Design Decisions. Consequently, ADRs emerge as a means to enhance Architectural Knowledge Management (AKM) within projects, ensuring that decisions made by various stakeholders over time can be understood within their historical context. Nevertheless, empirical evidence indicates a relatively low adoption rate of ADRs in open-source projects, despite an upward trend in recent years. 

The barriers to successful adoption of ADR are multifaceted: notably, the primary obstacles include the significant effort required to capture Architectural Knowledge (AK) and a lack of clarity regarding which specific AK elements warrant documentation \cite{Buchgeher2023}. Recent studies have explored the use of LLM-based approaches to automate ADR generation. There is evidence in literature about the feasibility of generating architectural decisions from textual context, while newer approaches combining Retrieval-Augmented Generation (RAG), few-shot learning, and fine-tuning improved both accuracy and practical viability for industrial use~\cite{Dhar2024,Dhar2025}. However, these approaches share a critical assumption that the input is already reasonably structured (e.g., a written document, a curated context, a codebase artifact). None of them were designed to handle raw, unstructured conversational data, where decisions are implicit, fragmented, and entangled with off-topic dialogue. Single-pass prompting (e.g., zero-shot, few-shot) struggles with noisy inputs, leading to attention degradation, unorganized outputs, and reduced accuracy due to irrelevant or excessive information~\cite{Gupta2026}.

To our knowledge, this is the first work to address ADR generation from raw, unstructured meeting transcriptions through an agentic, self-correcting workflow. We propose GADR, a multi-agent system that extracts, critiques, and refines architectural decisions from meeting transcripts, generating Nygard-formatted ADR drafts validated by both expert architects and the development teams themselves. If any recorded meeting can be turned into a reviewable ADR draft, with humans in the loop to validate the output, the primary adoption barrier for AKM lowers, and architectural knowledge can be captured at the moment it is created rather than reconstructed from memory. We present GADR as a proof of concept of an idea we believe is worth the research community's attention.

%Therefore, this research investigates the efficiency of LLMs in ADD extraction and ADR generation from student meeting transcriptions. Through an assessment across agentic, few-shot, and zero-shot approaches, the study shows how the use of real meetings as input can optimize the documentation workflow. This approach reduces the gap between verbal decision-making and formal recording, establishing a foundation for future integration with audio capture technologies and AKM automation.

\section{Background And Related Work}

\subsection{Difficulties In Identifying Design Decisions}

When conducted by students or novice architects, the architectural decision-making process is often highly fragmented and non-linear. Less experienced developers frequently face the complex challenge of transitioning from the problem to the solution space~ \cite{Capilla2020, van2010naive}. This difficulty stems from challenges in identifying suitable concepts, comparing technological alternatives, or building consensus \cite{Capilla2020}. Architectural decisions are critical to system structure and behavior, yet decision-making is often ad-hoc and intuition-driven, highlighting the need for methods and tools that support systematic reasoning and explicit decision documentation~\cite{van2010naive}.

Empirical studies reveal that unexperienced software engineers are heavily influenced by human and operational factors. For instance, it is highly common for them to select a technology, process, or approach with which they already have previous experience, rather than exploring new alternatives~\cite{Borowa2023}. When this prior knowledge does not align with the optimal solution for a specific problem, it can lead to misguided choices: as a result, the project meetings, and their transcriptions, become inherently unstructured and noisy, where concepts become entangled and ideas are easily lost. In such scenarios, the underlying decisions ultimately remain implicit, even to the participants engaged in the discussion.

As a way to address that gap, Architecture Decision Records (ADRs) emerged as lightweight artifacts for documenting decisions, context, rationale, and consequences~\cite{Nygard2011,keeling2017design}. However, their adoption still faces practical barriers, motivating recent LLM-based approaches for ADR generation. Zero-shot, few-shot, and fine-tuning strategies can synthesize ADR content from textual context~\cite{Dhar2024}. This process can be improved through RAG, few-shot learning, and fine-tuning~\cite{Dhar2025}. Nevertheless, these approaches remain challenged by long and noisy conversational inputs: excessive or poorly selected context can distract the model and reduce generation accuracy~\cite{Gupta2026}, which motivates investigating a multi-agent approach for extracting architectural decisions from meeting transcriptions.

\subsection{Knowledge Extraction Process From Text}
To bypass limitations in processing large, unstructured inputs, LLM orchestration has shifted from single-pass linear calls toward graph-based architectures and multi-agent systems. Prior work such as MetaGPT demonstrates that complex software engineering tasks can be decomposed into specialized agent roles coordinated through a structured workflow, improving the consistency of intermediate artifacts and reducing ambiguity in collaborative generation tasks \cite{Hong2024}. The use of traditional models often relies on linear workflows or simple recursive loops, which are limited when handling complex scenarios that require parallel execution, decision revision, or dynamic task routing \cite{Pelluru2025}. A graph-based orchestration approach, such as LangGraph, operationalizes this multi-agent decomposition by allowing developers to model the workflow as a finite-state machine (FSM), rather than submitting an entire transcription within a single prompt and expecting the model to extract and format decisions all at once. Within this framework, each node represents a specific step, edges determine the next node to be executed, and the state acts as a shared memory updated as execution progresses, thereby modularizing a large, complex problem.

Wang et al. explain that precise execution control derived from a unified state object enables the creation of fully adaptable workflows, whether sequential, hierarchical, or involving multiple collaborative agents \cite{Wang2025}. In the literature, multi-agent design patterns such as Planner-Executor (where one agent decomposes the task and another executes it) or Critic-Reviewer (where the output is validated step-by-step) demonstrate how modularization reduces hallucinations and improves factual accuracy in generated content \cite{Pelluru2025}. Therefore, adopting this architecture mitigates model attention degradation when processing large noisy inputs while ensures greater processing control by providing an auditable tool.

\section{Proposed Approach}
This section details the architecture of GADR (pronounced gather), a tool that automatically extracts architectural decisions from meeting transcriptions and generates structured ADR drafts. GADR moves away from linear single-pass pipelines by adopting an agentic architecture structured as a finite-state machine, augmented with Retrieval-Augmented Generation (RAG).

\subsection{Graph-State Formalization}
The control flow across the multi-agent framework is coordinated by a centralized, mutable state ($\mathbf{S}$). Within the LangGraph framework, using a shared state allows the system to dynamically update the context as node logic increases in complexity, enabling advanced routing and conditional looping capabilities. Rather than chaining sequential prompt calls, GADR coordinates its agents through a shared mutable state, making each reasoning step auditable and each transition explicit, which is a property that single-pass approaches cannot offer. Formally, this state is defined as a tuple that accumulates contextual variables generated across the graph nodes:
\begin{displaymath}
  S = (\tau_{orig}, \tau_{en}, \lambda, \Delta, \Gamma, \iota, \alpha)
\end{displaymath}
Each component plays a specific role in orchestrating the processed data stream:

    \noindent \textbf{transcript\_orig ($\tau_{orig}$):} The noisy, unstructured input representing the raw transcript of the meeting where the architecture was discussed.
    
    \noindent \textbf{transcript\_en ($\tau_{en}$):} The normalized representation of the transcript in English. This reduces tokenization biases and performance asymmetries across Large Language Models, ensuring that semantic processing occurs entirely in English.
    
    \noindent \textbf{language ($\lambda$): }A categorical control variable produced by the language detection module. The value of this variable dictates conditional routing to a translation node prior to information extraction.
    
    \noindent \textbf{decisions ($\Delta$):} A set of candidate values containing tuples structured as $\langle id, summary, textual\_context \rangle$. These represent preliminary ADR intentions extracted from the dialogue, which can be iteratively refined.
    
    \noindent \textbf{critique ($\Gamma$): }A qualitative feedback vector identifying extraction or architectural deficiencies (e.g., hallucinations, lack of justification, or omitted trade-offs). This vector functions as the grounding mechanism that drives the refinement node.
    
    \noindent \textbf{iteration ($\iota$):} A scalar counter that tracks the execution cycles of the self-reflection loop. It ensures algorithmic convergence, prevents infinite execution loops, and guarantees that the system reaches a terminal state within an acceptable time boundary.

    \noindent \textbf{adrs ($\alpha$): }Stores the final generated artifacts, formatted strictly according to Nygard's ADR template.

\subsection{Nodes Specification}
\begin{figure}[ht]
  \centering
  \includegraphics[width=.85\linewidth]{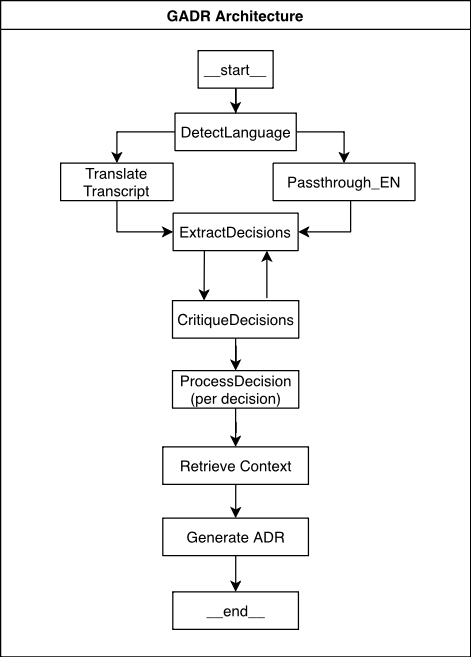}
  \caption{Informal overview of GADR's processing workflow and state transitions.}
  \label{fig:local-architecture}
  \Description{Informal control-flow diagram in which boxes represent GADR processing nodes and directed arrows represent state transitions, including a conditional language branch and a critique-refinement loop.}
\end{figure}

The workflow is decomposed into nodes that read and update the shared state $S$. Each node performs a bounded operation, such as an LLM call, critique step, retrieval operation, or formatting routine. The local architecture is shown in~\autoref{fig:local-architecture}, combining local modules and external services used by the agent.

At runtime, GADR coordinates local components with external services. Local modules manage transcript preprocessing, state transitions, prompt orchestration, ADR persistence, and retrieval over ChromaDB. External calls are used when the workflow requires LLM reasoning through the Gemini API, configured with the Gemini 3.1 Pro model, and web search through the Tavily API. This separation keeps project-specific artifacts and vector retrieval local while delegating generation and current technical search to specialized services.

\noindent\textbf{Input Language Alignment Module.} Because LLM reasoning and tokenization are often more efficient in English, especially when compared with non-English inputs \cite{schut2025}, and prior evaluations report better factual accuracy when prompting in English \cite{Rohera2025}, the workflow first normalizes the transcript language. The DetectLanguage node classifies a representative sample of the transcript, formalized as $T_{orig}[start:start+1000]$. If the transcript is classified as non-English ($\lambda="other"$), the Translate node produces an English version; otherwise, Passthrough\_EN copies the original text to the normalized transcript field. This keeps downstream extraction consistent while avoiding unnecessary translation.

\noindent\textbf{Architectural Decision Extraction Node (ExtractDecisions.)} This node extracts candidate architectural decisions from the normalized transcript. In the first iteration $(\iota=1)$, the prompt asks the model to act as a senior software architect and identify decisions, rationale, and rejected alternatives that affect the project architecture. In later iterations $(\iota>1)$, the node uses the critique vector to refine the previous output by merging overlapping decisions, splitting tangled concerns, and removing noise.

\noindent\textbf{Critical Evaluation Module (CritiqueDecisions).} This node reviews the provisional decisions before ADR generation, reducing hallucinations, overlaps, and inconsistent granularity. Its feedback is organized in categories (~\autoref{table:cat}) that guide the refinement loop.

\begin{table}\footnotesize
  \caption{Decision critique categories.}
  \label{table:cat}
  \begin{tabularx}{\linewidth}{l X}
    \toprule
    Category & Description\\
    \midrule
    Overlap              & Merges decisions that represent the same architectural strategy. \\
    Fragmentation        & Detects decisions that are too small to justify a separate ADR. \\
    Missing              & Finds relevant choices or trade-offs omitted from extraction. \\
    Noise                & Removes procedural decisions without architectural impact. \\
    Scope Duplicate      & Removes items that only repeat broader decisions. \\
  \bottomrule
\end{tabularx}
\end{table}

\noindent\textbf{RAG Enrichment and Final Formatting Node (ProcessDecision).} It enriches each accepted decision, formatting it as a Nygard-style ADR. It performs hybrid retrieval in a local ChromaDB knowledge base and web search results, re-ranks the retrieved content to keep only the most relevant technical facts, and generates a Markdown ADR with Context, Decision, Considered Options, and Consequences. This step improves architectural explanation, but its retrieved content remains subject to human review.

\section{Methodology And Setup}

Our primary goal is to analyze the effectiveness of the proposed multi-agent system in extracting and documenting Architectural Decision Records (ADRs) from raw meeting transcripts. %from the perspective of software engineering students.

\subsection{Engineering Research}
This study follows an Engineering Research methodology, which, according to SIGSOFT's SE Empirical Standards~\cite{ralph2021acm}, consists of ``research that invents and evaluates technological artifacts''. In this case, a novel software engineering artifact was designed, implemented, and empirically evaluated in an educational software architecture setting. The evaluation adopts an exploratory mixed-method design combining: (i) comparative analysis against traditional prompting strategies (zero-shot and few-shot), (ii) expert review conducted by senior software architecture evaluators, and (iii) participant-centered evaluation using questionnaires and qualitative feedback from students. Given the early stage of this research, the evaluation is intentionally scoped as a feasibility probe rather than a definitive comparison.

%The goal of the study is not to establish definitive performance claims, but rather to investigate the feasibility, perceived usefulness, and comparative behavior of an agentic workflow for ADR generation from noisy meeting transcriptions.

This research was reviewed and approved by the university's Research Ethics Committee\footnote{The CAAE (Certificado de Apresentação para Apreciação Ética) number is omitted to comply with the anonymous review.}. All procedures followed national ethical guidelines, with informed consent obtained prior to data collection. Students were informed that participation in this study would not interfere in their grades.

To systematically achieve this goal and evaluate the proposed approach against traditional prompting techniques, this study addresses the following Research Questions (RQs):

    \noindent \textbf{RQ1 (Extraction Effectiveness):} To what extent does the agent identify and extract plausible architectural decisions discussed in meetings?
    
    \noindent \textbf{RQ2 (Artifact Quality):} Do the generated ADRs possess sufficient clarity and completeness to be utilized as practical documentation artifacts?
    
    \noindent \textbf{RQ3 (Recall Utility):} To what extent do generated ADRs help users recall architectural decisions discussed in previous meetings?
    
    \noindent \textbf{RQ4 (Comparative Analysis):} What differences can be observed in comparison to traditional zero-shot and few-shot prompting strategies when extracting ADRs from noisy conversational data?

\subsection{Study Context And Participants}
The evaluation combines curated baseline of existing ADRs with real-world conversational data:

\noindent \textbf{Reference Dataset (MSR Study):} To establish a ground truth for formatting and technical validation, this study uses the dataset provided by Buchgeher et al., comprising ADRs extracted from open-source repositories available at the time of writing \cite{Buchgeher2023}. This dataset serves two functions: it acts as the primary knowledge base for the RAG module and provides the "gold standard" semantic examples utilized in the few-shot baseline evaluation.

\noindent \textbf{Transcription Dataset:} The primary input data consists of raw audio transcripts recorded during software design meetings of four teams of undergraduate students and one team of senior researchers in an architecture kick-off meeting of an R\&D project (\autoref{tab:transcripts}). These transcripts represent unstructured and noisy environments, characterized by informal language, transcription errors, overlapping dialogue, and fragmented rationale, providing a robust stress test for the extraction method.

\begin{table}[ht]\footnotesize
\caption{Characterization of the five groups (G) meetings.}
\label{tab:transcripts}
\centering
\begin{tabular}{p{0.4cm} p{1.3cm} p{2.5cm} p{3cm}}
\toprule
\textbf{G} & \textbf{Team} & \textbf{Context} & \textbf{Decision Themes} \\
\midrule
1 & 6 Students & Online FPS game & Networking model, server authority, latency handling, event broker, object pooling \\
\addlinespace
2 & 7 Students & Urban mobility/logistics & Event-driven architecture, message broker, external-event simulation, integration facade \\
\addlinespace
3 & 7 Students & Smart gym platform & AWS/EKS deployment, autoscaling, brokers, Redis/BullMQ, offline synchronization \\
\addlinespace
4 & 4 Students & AI/multi-agent system & Layered architecture, hexagonal variant, event-driven trade-offs, future microservices \\
\addlinespace
5 & 4 Senior Researchers & Health R\&D platform & Offline support, Gov.br authentication, RNDS integration, C4/ADR documentation \\
\bottomrule
\end{tabular}
\end{table}

\subsection{Evaluation Procedure}
To evaluate the effectiveness of the proposed solution, this study conducts a comparative analysis between the multi-agent system and two established prompting approaches. The transcripts were processed uniformly across three different setups, as follows.

    \noindent \textbf{- Zero-shot Prompting:} The raw transcript is submitted to the LLM accompanied by a direct system prompt to generate the ADR following the Nygard standard. This scenario tests the model's raw inference capability over noisy data without contextual examples.
    
    \noindent \textbf{- Few-shot Prompting:} The prompt context is enriched with static examples of Architectural Decisions extracted from the reference MSR dataset. This evaluates whether static examples are sufficient to guide the model's extraction and formatting logic.
    
    \noindent \textbf{- Multi-agent Workflow:} The transcript is processed through the multi-node directed graph detailed in Section 3.

This comparison is key because traditional single-pass approaches, such as zero-shot and few-shot, often suffer from attention degradation when processing lengthy, unstructured texts. This limitation frequently leads the model to hallucinate or ignore crucial rationale. The agentic approach aims to quantify the gains in cohesion and accuracy provided by a step-by-step resolution.

\subsection{Evaluation Strategy}

Because automatic Natural Language Processing metrics primarily capture textual similarity rather than the practical quality of architectural documentation, this study prioritizes human-centered evaluation \cite{Gupta2026,Dhar2025}. The generated ADRs were assessed through an anonymous online questionnaire distributed to the five teams that provided meeting transcripts, restricted to participants who gave informed consent. The students had varying levels of software development experience and evaluated ADRs generated for the projects they developed in the software architecture course.

The questionnaire collected contextual information using five-point Likert scales to assess agreement with extracted decisions, clarity, completeness, and perceived educational value, with an open-ended field for justification. To complement participant feedback, four senior software architecture reviewed the five transcripts to identify core decisions discussed by each team. The expert-identified decisions are a ground-truth approximation against which the agent's ADRs were compared in number and content.

%The expert baseline is compared only with the outputs produced by the proposed agentic approach. We compare the number of architectural decisions identified by the senior evaluators with the number of ADRs generated by the agent, as well as the main decision documented in each ADR. This mapping provides evidence of whether the more deterministic agentic workflow can approximate an acceptable number of decisions and, more importantly, whether it captures the same core architectural decisions identified by expert judgment. Thus, the evaluation combines participant perception, qualitative feedback, and expert comparison, reducing the risk of relying only on subjective student impressions.

\section{Results}

\begin{table*}[ht]\footnotesize
\centering
\caption{Summary of evaluation results across three complementary perspectives.}
\label{tab:evaluation-summary}
\begin{tabular}{p{2.8cm} p{2.8cm} p{5.5cm} p{4.5cm}}
\toprule
\textbf{Data Source} & \textbf{N} & \textbf{Key Finding} & \textbf{Main Limitation} \\
\midrule
Comparative (Agentic vs.\ Zero/Few-shot) 
& 5 transcripts $\times$ 3 rounds 
& Agentic RAG produced lower variance and richer ADRs; zero-shot was least stable 
& Verbosity increases review cost \\
\addlinespace
Participant Forms 
& 15 respondents, 55 decision-level evaluations 
& 52/55 agreed with extracted decision; 51/55 found ADRs clear and complete 
& Students evaluated their own projects, introducing familiarity bias \\
\addlinespace
Expert Evaluation 
& 4 evaluators, 23 reference decisions 
& Agent covered {\raise.17ex\hbox{$\scriptstyle\sim$}}18/23 expert-identified decisions 
& Agent occasionally treated tentative discussions as accepted decisions \\
\bottomrule
\end{tabular}
\end{table*}

\subsection{Agentic Versus Prompt-Based Generation}

We report preliminary results from a multi-round evaluation of three ADR generation strategies: zero-shot prompting, few-shot prompting, and the proposed Agentic RAG workflow. Five real software architecture meeting transcripts were independently processed three times for each approach. Because LLM outputs are non-deterministic, the comparison focuses on variance in the number of generated ADRs, adherence to naming and formatting conventions, output length, and evidence of hallucination or contextual noise.

\begin{comment}
\begin{table*}[t]\footnotesize
\centering
\caption{Preliminary Comparison Across Three Execution Rounds}
\label{tab:preliminary-results}
\begin{tabularx}{\textwidth}{lXXX}
\toprule
Approach & Determinism & Format & Output Behavior \\
\midrule
Zero-shot & High ADR-count variance across rounds. & Weak; irregular titles and sections. & Shorter outputs, but more superficial and noisy. \\
Few-shot & Medium variance, with better decision grouping. & Stronger adherence to the Nygard structure. & Balanced length, but still occasionally splits related decisions. \\
Agentic RAG & Lower variance in core decisions. & Strong file naming and standardized Nygard structure. & Longest outputs, with richer context and higher review cost. \\
\bottomrule
\end{tabularx}
\end{table*}
\end{comment}

\begin{figure*}
    \centering
    \includegraphics[width=1\linewidth]{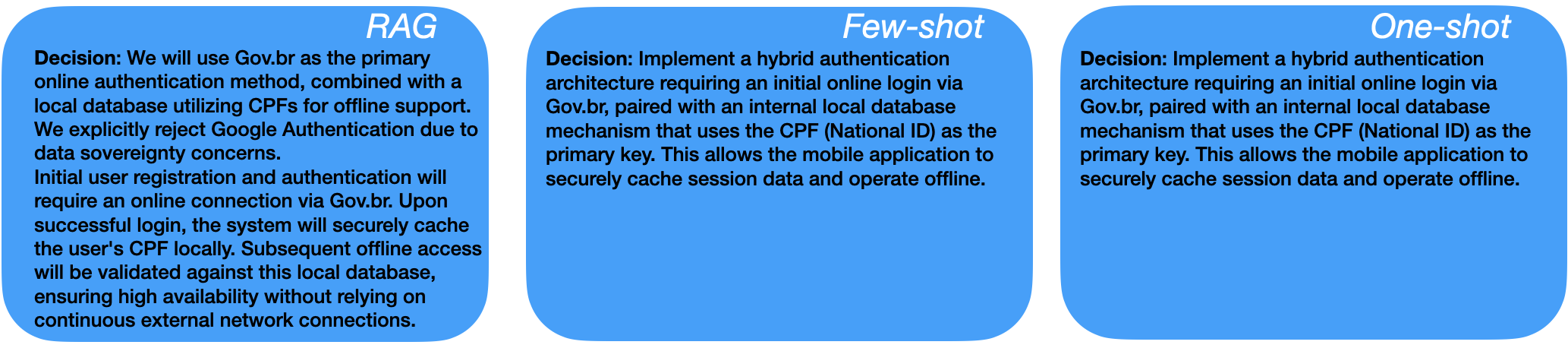}
    \caption{Excerpts of an ADR generated from the same transcription (Group 5) but in different approaches.}
    \label{fig:adr-examples}
    \Description{Side-by-side excerpts of ADRs generated from the Group 5 transcript using different approaches, illustrating differences in structure, detail, and decision documentation.}
\end{figure*}

The static prompting baselines were more sensitive to round-to-round variation. Zero-shot outputs were the least stable, sometimes merging unrelated decisions or extracting passing comments as ADRs. Few-shot prompting improved structure and formatting by following the provided examples, but still varied in how it grouped decisions. In contrast, the Agentic RAG produced more stable sets of ADRs (e.g., \autoref{fig:adr-examples}). Its critique stage reviewed candidate decisions before document generation, reducing overlaps and fragmentation.

The main advantage of the Agentic RAG approach was not only structural consistency, but also the production of richer architectural explanations. This verbosity is useful in the educational context of this study, since novice students often discuss decisions informally and do not explicitly articulate trade-offs, alternatives, or long-term consequences. By enriching ADRs with architectural terminology and external context, the agent can transform a sparse meeting transcript into a more informative learning artifact.

However, this enrichment also creates a risk of contextual contamination. In the FPS game case, the agent added to the ADR context that the network architecture should support ``500 concurrent connections and 5000+ Daily Active Users (DAU)'', although these metrics were not present in the transcript. Traceability inspection showed that this fragment came from a retrieved external ADR, \texttt{0005-discovery-protocol.md}, rather than from the meeting itself. This illustrates how retrieval can lead to hallucinations or unsupported statements when semantically similar ADRs come from unrelated projects. At the same time, it also shows the potential value of using a project's own historical ADRs as retrieval context: previous decisions, constraints, insights, or metrics can be brought back into new ADRs and revisited during architectural reasoning. Therefore, the current results indicate a trade-off: Agentic RAG improves determinism, formatting, and pedagogical depth, but its outputs require human review to distinguish faithful extraction from retrieved-but-unrelated context.

\subsection{Participant Evaluation Through Forms}

The participant evaluation included 15 respondents from the five student groups, producing 55 decision-level evaluations because each participant assessed multiple generated ADRs. Overall, the responses indicate a positive perception of the agentic ADRs: 52 of 55 evaluations agreed or strongly agreed with the extracted decision, 51 of 55 indicated that the suggestions helped participants remember previous decisions, 51 of 55 considered the ADRs clear and complete, and 49 of 55 agreed that the artifacts contributed to learning. The open-ended responses reinforce this result, with recurring comments that the ADRs captured the discussion, reflected industry-like reasoning, and helped participants recover the context and rationale behind previous decisions. At the same time, these qualitative answers reveal important limitations: in Group 2, one participant noted that a Python script was discussed as a possibility but not definitively decided; in Group 1, disagreement around the MVC and Event Broker ADR suggests that the agent may impose a stronger architectural framing than the group explicitly adopted; and in the Group 5 case, neutral feedback indicated that some decisions required clearer contextualization. Thus, the form results support the usefulness and perceived clarity of the generated ADRs, while also showing the need for human review before treating them as final documentation.

\subsection{Senior Developers Evaluation}
The expert evaluation was conducted not to certify the agent's output as correct, but to assess whether the agentic workflow approximates the decisions a trained architect would identify from the same transcript. Across the five transcripts, the seniors listed 23 reference architectural decisions, while the agent generated 17 ADRs. Using a decision-level mapping, approximately 18 of the 23 senior decisions were covered by at least one generated ADR, suggesting that the agent captured most of the central architectural concerns discussed in the meetings. The strongest alignments occurred when decisions were explicit in the transcripts, such as event-driven architecture and message brokers in Group 2, AWS/EKS/BullMQ/Redis in Group 3, and offline support, Gov.br authentication, and integration concerns in the Group 5 project.

The comparison also exposed relevant failure modes. First, the agent sometimes transformed tentative discussions into accepted decisions, as observed with the MVC and Event Broker ADR in Group 1 and the treatment of Lambda as a rejected alternative in Group 3. Second, it occasionally introduced unsupported quantitative or statistical claims, such as ``5000+ Daily Active Users'' and ``93\% industry adoption'', which were not present in the transcripts. Third, some lower-granularity decisions identified by seniors, such as object pooling in Group 1 or load balancing and auto-scaling in Group 3, were omitted or only mentioned indirectly. Finally, the analysis showed that the expert baseline itself may be incomplete in some cases: for Group 4 and Group 5, the transcripts contained evidence supporting decisions generated by the agent but not listed by the senior evaluator. Overall, the expert comparison indicates that the agentic approach approximates the number and content of human-identified decisions reasonably well, but still requires expert review to correct overstatements, unsupported details, and mismatches in decision status.

\section{Discussion}
The results indicate that the proposed agentic workflow is better understood as a decision-support mechanism than as a fully autonomous replacement for human architectural documentation. Compared with zero-shot or few-shot prompting, the agentic approach produced more stable outputs and stronger adherence to the ADR structure, suggesting that decomposing the task into extraction, critique, refinement, retrieval, and formatting reduces some of the instability caused by processing noisy transcripts in a single prompt. This backs our main assumption: the extraction of architectural decisions from meetings benefits from explicit orchestration.

Considering the research questions, the results provide preliminary answers to all four. For \textbf{RQ1}, the agent covered approximately 18 of the 23 expert-identified decisions, suggesting effective extraction of central concerns; however, omissions and confusion between tentative and accepted decisions prevent interpreting this result as full accuracy. For \textbf{RQ2}, 51 of 55 evaluations considered the ADRs clear and complete, supporting their use as reviewable drafts rather than final records because retrieved details were not always transcript-grounded. For \textbf{RQ3}, 51 of 55 evaluations indicated that the ADRs helped participants recall prior decisions. This supports recall utility in the studied setting, although participants' involvement in the projects may have influenced this perception; their reported learning benefit is therefore treated as a complementary finding rather than part of RQ3. For \textbf{RQ4}, the three-round comparison showed lower variation and stronger structural adherence for the agentic workflow than for zero-shot and few-shot prompting. These gains came with greater verbosity and review effort, indicating a trade-off rather than an unconditional advantage.

The participant evaluation reinforces the value of this approach in an educational setting. Students not only recognized most extracted decisions, but also reported that the ADRs helped them remember previous discussions and better understand architectural documentation. This pedagogical benefit is relevant because novice architects often discuss decisions informally, without clearly separating context, alternatives, consequences, and rationale. By converting these discussions into structured ADRs, the tool can help students revisit their own reasoning and learn how architectural trade-offs are documented in practice.

However, the results also show that richer documentation is not necessarily more faithful documentation. The Agentic RAG workflow generated more detailed ADRs, but some details came from retrieved external context rather than from the transcript itself. Unsupported metrics, statistical claims, and stronger-than-intended architectural framing illustrate how retrieval can improve explanation while also introducing contextual contamination. Therefore, enrichment should be treated as useful but provisional, requiring traceability and human validation.

Expert review complemented student feedback by checking transcript grounding. GADR captured most senior-identified decisions and central architectural concerns, but sometimes promoted tentative ideas to accepted decisions, omitted fine-grained concerns, or grouped related choices into broader ADRs. Future versions should expose decision status and granularity, enabling users to inspect, split, merge, or reclassify candidates before documentation.

Overall, GADR appears most useful as a semi-automated documentation assistant. It reduces the effort required to transform informal meetings into readable ADR drafts and provides educational value for students learning architectural reasoning. Nevertheless, the generated artifacts should remain reviewable drafts: project stakeholders must still correct unsupported extrapolations, confirm decision status, and decide which ADRs should become part of the official architectural record.

\section{Threats To Validity}
Threats arise from the use of LLM-based and RAG-based generation. Although the agentic workflow improved structure and stability when compared with single-pass prompting, it sometimes transformed tentative discussions into accepted decisions, omitted fine-grained concerns, or introduced unsupported details from retrieved external context. In addition, verbose ADRs may be useful as learning artifacts for novice architects, but verbosity can also obscure which information was actually discussed in the meeting and increase the review effort. Therefore, the generated ADRs should be interpreted as reviewable drafts rather than authoritative documentation, and unsupported additions should be treated as potential hallucinations instead of automatic improvements.
This study is also subject to threats related to its exploratory and mostly educational setting. The evaluation was conducted with five student groups and a limited number of transcripts, which restricts the generalization of the findings to industrial contexts. Moreover, students evaluated ADRs from projects in which they participated, so their perception may include implicit knowledge not fully present in the transcripts. This is especially relevant for the reported pedagogical gain: students perceived the ADRs as useful for remembering decisions and learning architectural documentation, but this benefit may also be influenced by their prior involvement in the projects and by the explanatory richness of the generated texts.

\section{Conclusion}
This paper presented an agentic approach for extracting architectural decisions from meeting transcriptions and generating ADR drafts. Building on the role of architectural decisions in software architecture \cite{Jansen2005} and on ADRs as lightweight documentation artifacts \cite{Nygard2011,Nogueira2026}, the approach addresses a known adoption barrier: the effort required to capture architectural knowledge in practice \cite{Buchgeher2023}. The preliminary evaluation indicates that, compared with zero-shot and few-shot prompting, the agentic workflow produced more stable and structured ADRs, captured most expert-identified decisions, and supported students in remembering and learning from their own architectural discussions.

At the same time, the results reinforce that LLM-based ADR generation should remain human-in-the-loop. Although prior work shows the promise of LLMs and RAG for architectural decision generation \cite{Dhar2024,Dhar2025}, this study suggests that if any recorded meeting can become a reviewable ADR draft, then the primary adoption barrier for AKM shifts in nature: from the effort required to write documentation to the lighter effort of reviewing it. We believe this shift is worth pursuing, and present GADR as a concrete first step.

%retrieved context and verbose explanations can introduce unsupported details. Future work will therefore focus on improving traceability between transcript fragments and generated ADR sections, making decision status explicit and compared to real written ADRs.

\section*{Artifact Availability}
The research artifacts supporting this study are archived on Zenodo at
\url{https://doi.org/10.5281/zenodo.21501559}. The project repository is also
available at \url{https://github.com/ldcss/GADR}.

\begin{acks}
For partially supporting this work, we would like to thank INES.IA (National Institute of Science and Technology for Software Engineering Based on and for Artificial Intelligence) www.ines.org.br, CNPq grant 408817/2024-0. 
\end{acks}
\bibliographystyle{ACM-Reference-Format}
\bibliography{references-base}

\end{document}